# Radio Signal Correlation at Frequency 32 MHz with Extensive Air Showers Parameters Using Yakutsk Array Data.

S. P. Knurenko[1, a], I. S. Petrov[1, b].

[1]Yu. G. Shafer Institute of Cosmophysical Research and Aeronomy SB RAS, Yakutsk

The paper present correlation of radio signal with air shower parameters: shower energy $E_0$ and depth of maximum $X_{max}$. It is shown that from radio emission measurements of air showers one can obtain individual showers parameters and mass composition of cosmic rays. We also derived generalized formula for calculating energy of the air showers.

## 1. INTRODUCTION

Study of cosmic rays properties by measuring the radio emission generated by charged particles of extensive air showers may be an alternative method to traditional methods that use large areas of the arrays. The arrays are consist hundreds and thousands of scintillation detectors for registration of charged particles, or consist detectors, recording emission generated by relativistic particles of EAS in the optical wavelength range. Such arrays are very costly because of a large amount of detectors and complex technical equipment. On the other hand, radio method is much cheaper and easier to operate with nearly 100% duty cycle. It is sufficient to have the antenna field and a simple radio receiver tuned to a given frequency. The main problem is to choose a noise free frequency range. For this purpose, in Yakutsk was installed and started radio array for EAS radio emission. The array consists crossed antennas oriented E - W and N - S. Air shower radio registration is conducted at a frequency of 32 MHz, free from industrial noise. Yakutsk Radio Array operates since 2008. Data obtained during those several seasons includes showers with energy above $10^{19}$ eV.

## 2. THEORETICAL BACKGROUND

Generation of radio emission mechanisms are well known and calculations [1] shows that the radio emission of EAS depends on the development of an electromagnetic cascade in the atmosphere, magnetic field near the level of observations, etc. That is, the radio emission is associated with the physics of air showers development: production of electrons and positrons along the way of particles in the atmosphere (longitudinal development of air shower) and total number of charged particles (EAS energy). During the calculations, formulas that reflect connection of radio pulse amplitude with air shower characteristics were obtained.

---

[a]s.p.knurenko@ikfia.sbras.ru  [b]igor.petrov@ikfia.sbras.ru

The electric field induced on the antennas of the radio arrays, according to [1], can be represented as follows:

$$\varepsilon_\nu = \sqrt{\frac{4\pi\nu^2\mu_0}{G_{(\theta,\phi,\nu)}c}\frac{1}{A_{ele(\nu)}R_{ADC}}}\frac{V_{ADC}}{\Delta\nu} \qquad (1)$$

It is also shown that the maximum amplitude of the radio pulse is proportional to the energy $A_{max}$ electromagnetic component of EAS with $E_{em} = c \cdot A_{max}$ (µV$^{-1}$ m GeV), where c - parameter of air shower energy.

From radio measurements data and calculations [1], we can determine the shower energy regardless of registration method and then compare two estimations.

In paper [1] performed calculations, which showed that the slope of radio emission LDF depends on the depth of maximum $X_{max}$. Slope of LDF determined as the ratio of the amplitudes of the radio pulse at flat and steep part of LDF. In this case, the distance of 175 m and 725 m from the shower axis were used. Analytically, this relationship can be expressed by the formula:

$$X_{Max} = a[\ln(b(A_{175\,m} / A_{725\,m}))]^c \qquad g/cm^2, \qquad (2)$$

Here a, b, c – constant coefficients, a = 856,1, b = 0,3149, c = 0,4340.

## 3. SHORT DESCRIPTION OF YAKUTSK RADIO ARRAY

First experiments to register radio pulses from air showers were have been made in 1987 – 1989 years at Yakutsk array. Were registered radio signal in 6250 showers with energy more than 10$^{17}$ eV, including some events with E0 ≥ 10$^{19}$ eV. In 2009, at Yakutsk array were installed 6 antennas (half-wave dipole) for radio emission registration. In 2009-2011 radio emission registration were conducted by 6 antennas placed at 300, 350 and 500 m from the array center. For optimal frequency choice background frequency spectrum from 1 to 200 MHz were analyzed. At frequencies, up to 20 MHz due to the presence of large natural radio noise primarily storm origin it is not possible to distinguish air showers pulses with sufficient efficiency. Therefore, it is advisable to select frequency above 20 MHz. Amplitude of galactic noises is much smaller decreases with frequency increasing comparing to storm noises and equals to 1.0 – 2.0 µV*m$^{-1}$ *MHz$^{-1}$ at 32 MHz. Thermal noises of antenna much smaller than galactic noises at frequencies up to 100 MHz and almost have no influence to measurements. Therefore, the most favorable for the measurements range at Yakutsk array should be considered narrow range of frequencies 30 – 40 MHz. For registration at frequency around 32 MHz with bandwidth 8 MHz array was installed. The array consists of antennas, amplifiers and registration device with data storage [4]. We used half-wave dipole lifted to λ/4 for registration radio emission from air showers. Crossed dipoles are oriented in an east - west (on the magnetic latitude), and north - south (along the magnetic meridian). Bandwidth is ±4MHz, sensitivity ~10μV (2 μV*m$^{-1}$ *MHz$^{-1}$), dynamic range is 50 dB.

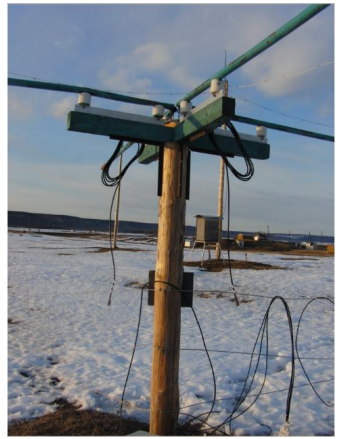

Fig. 1. Design of the receiving antenna

Receiving channels based on the principle of direct signal amplification and subsequent detection. Antenna amplifiers are placed into special containers and are located directly at the antenna. The main paths are based on the cascade amplification circuit with mismatched contours.

The array currently consists two set of antennas, distance between them is 500 m. First set consists 8 antennas, second 4 antennas. The antenna is a perpendicularly crossed dipoles, one to direction W-E, other N-S and lifted to the height 1.5 m above the ground. Each set is connected to an industrial computer. Directly under the antennas are electronic devices and matching amplifiers and shielding grid on the ground. Testing and calibration of measurement including the antennas is done once in a measurements season using a broadband high-frequency generator and a control PC. Registration of radio emission is triggered by one of two event triggers (masters) from Yakutsk array. Location of antenna sets is shown in Fig. 2.

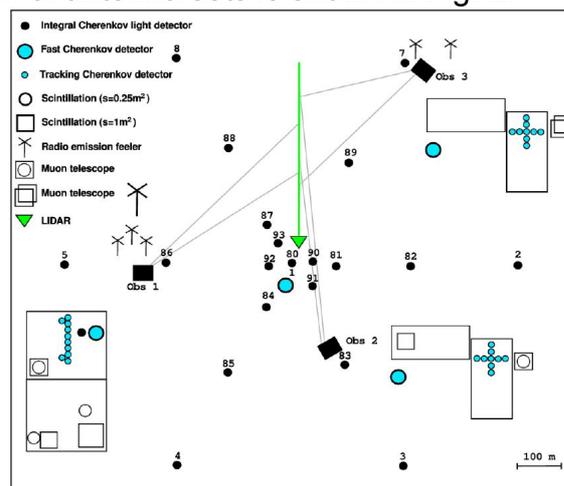

Fig. 2. Arrangement of observation stations on the Small Cherenkov array.

Air shower events is triggered by one of two triggers (master signal). The first is the main Yakutsk array trigger, registers showers in area 12 km$^2$ with energy more than $10^{17}$ eV. The second Small Cherenkov array, registers showers in area 1 km$^2$ with energy $10^{15} - 5 \cdot 10^{17}$ eV. EAS radio emission registers in the case if one of two trigger signals has been arrived. This trigger signal is the control signal for the registration of radio pulses received from antenna cable to the ADC. The data collection system is based on an industrial computer and is able to recording signal synchronously from multiple antennas simultaneously. We used fast 8-bit ADC, which provided recording prehistory (before the trigger arrival) for 25 ms and history (after the arrival of the trigger) for 15 ms (Fig. 3).

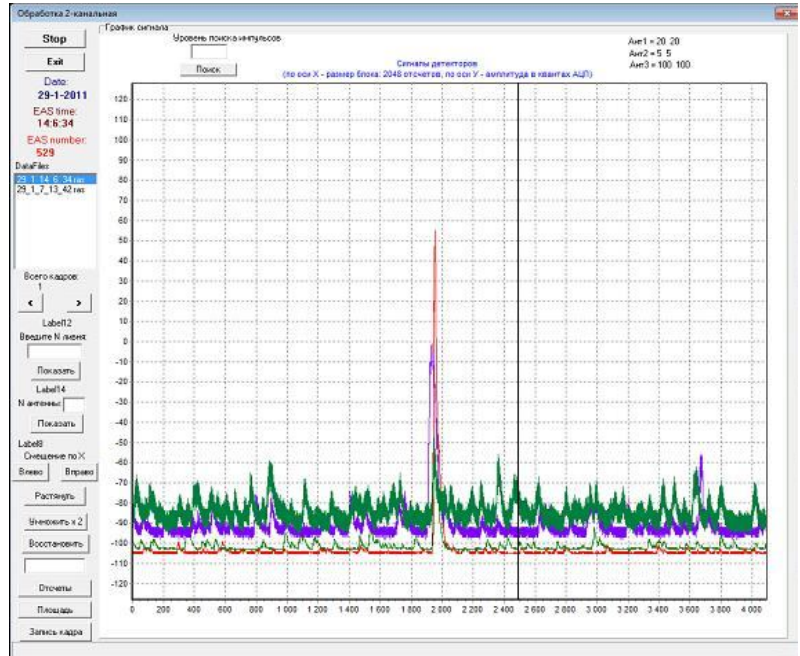

Fig. 3. Radio emission pulses from 4 antennas.

In Fig. 3, OX axis 40 μs time scale, OY axis signal amplitude in ADC samples. Vertical line is the moment of the trigger signal arrival.

## 4. RESULTS

As it shown in Fig. 4 average slope of the LDF varies with distance. At large distances from the shower axis, radio signal attenuates slowly.

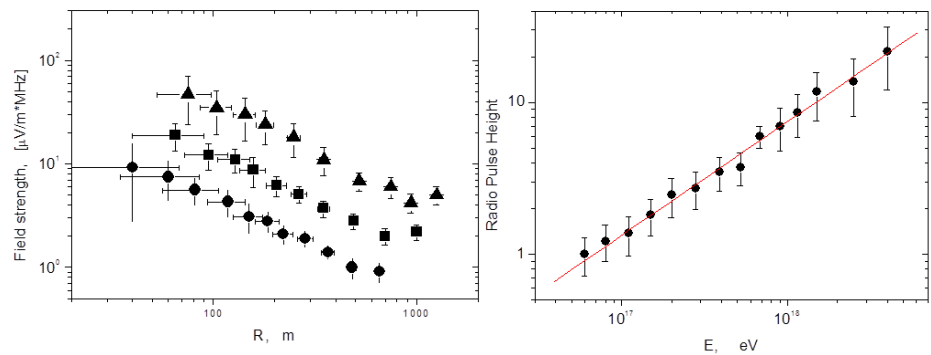

Fig. 4. Average LDF of radio emission at 32 MHz in showers with energies $1{,}73 \cdot 10^{17}$ eV, $4{,}38 \cdot 10^{17}$ eV and $1{,}32 \cdot 10^{18}$ eV (left). Dependence of the maximum amplitude of the radio emission pulse from the shower energy (right).

From formulas (1) and (2), and measurements of the radio pulse we empirically found a correlation of radio emission amplitude with shower energy $E_{em}$, which was determined by the flux of Cherenkov light, and the shape of the LDF of the signal we found a correlation with $X_{max}$.

For this analysis, we selected events that consist simultaneously measured charged particles, muons and Cherenkov light. This allowed us to establish a connection between radio emission and shower characteristics – energy $E_0$ and depth of maximum $X_{max}$.

The only thing that prevents us from a full comparison – it is a small statistics. However, some results were obtained, for example, the correlation of the shower energy, which is obtained by using a full flux of EAS Cherenkov light with a maximum amplitude of the radio signal. This result is shown in Fig. 5 and approximation is given by formula (3).

$$\varepsilon_{EW} = (1{,}3 \pm 0{,}3)(E_0 / 10^{17} \text{ eV})^{(0{,}99 \pm 0{,}04)} \quad [\mu V/m/MHz] \quad (3)$$

where, $\theta$ - zenith angle, R - the distance to the antenna, $E_p$ - energy of the primary particle. Generalized formula (4) makes it possible to determine the energy of individual shower coming with zenith angle $\theta$, the amplitude of the radio signal registered at a distance of 350 m from the shower axis:

$$\varepsilon = (15\pm1)\,(1-\cos\theta)^{1{,}16\pm0.05} \cdot \exp(-R/(350\pm25.41)) \cdot (E_p/10^{17})^{0.99\pm0.04} \quad (4)$$

In Fig. 6: solid curve – calculations by formula (2); dots – Yakutsk experimental data. Fiducial values of amplitude were taken at same distances as [1]. Fig. 5b shows that the experimental data are in a good agreement with calculations.

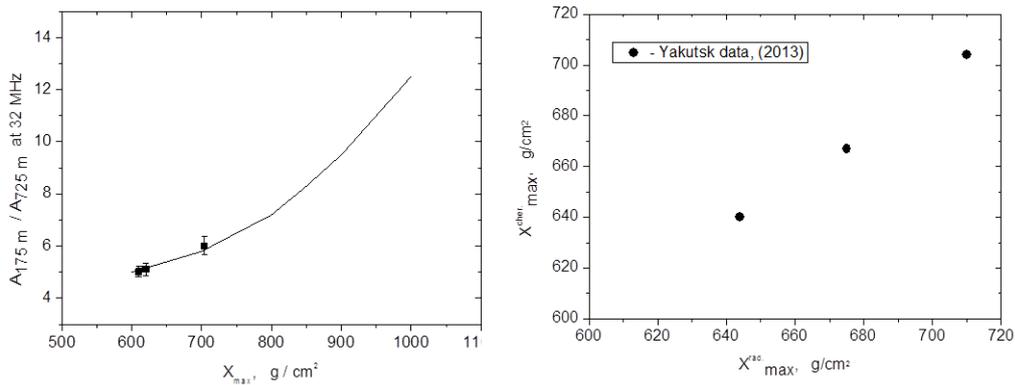

Figure 5a. Ratio of the amplitudes of radio emission in the steep and flat part of LDF with the depth of maximum distribution $X_{max}$ of individual shower. Curve is a calculation by formula (2) with coefficients a = 856.1, b = 0.3149, c = 0.4340. Figure 5b. Depth of maximum correlation obtained from measurements of Cherenkov light $X_{opt.max}$ and from measurements of the LDF of the radio emission using the formula (2) $X_{rad.max}$.

In paper [2] calculations show that the shape of depth of maximum $X_{max}$ at fixed energy represents the contribution of different groups of nuclei and it can be used to analyze the mass composition of cosmic rays.

## 5. CONCLUSION

Yakutsk array data show that: a) there is a correlation between measured amplitude of the radio signal with shower energy; b) depth of maximum determined from measurements of Cherenkov detectors $X_{opt.max}$ and calculated by formula (2) $X_{radmax}$ are in a good agreement with adjusting for noise contribution in radio [3].